# Using Causal Threads to Explain Changes in a Dynamic System

Robert B. Allen [0000-0002-4059-2587]
New York, NY, USA

**Abstract:** We explore developing rich semantic models of systems. Specifically, we consider structured causal explanations about state changes in those systems. Essentially, we are developing process-based dynamic knowledge graphs. As an example, we construct a model of the causal threads for geological changes proposed by the Snowball Earth theory. Further, we describe an early prototype of a graphical interface to present the explanations. Unlike statistical approaches to summarization and explanation such as Large Language Models (LLMs), our approach of direct representation can be inspected and verified directly.

## 1  INTRODUCTION

We envision interactive digital libraries that are built on knowledge bases rather than repositories of text. In this work, we focus on qualitative semantic models of systems. Systems may be defined as collections of objects that interact in stable and predictable ways over time. In (Allen, in preparation) we explore the description of systems at equilibrium. Those are systems in which none of the major states are changing (although the states of subsystems may be changing). In this companion paper, we consider qualitative descriptions of causal processes for dynamic systems with a higher-level implementation of the transitions. Our goal is to provide an explanation for users of how changes in one part of the system lead to system state changes.

In Section 2, we review our approach to structured descriptions and then consider discourse and causal explanations for dynamic models. In Section 3, we apply our approach to structured description to the Snowball Earth theory and introduce a visualization interface to support user interaction. Section 4 describes features to be explored in future work.

## 2  SEMANTIC MODELS

### 2.1  Structured Descriptions

In an approach we call direct representation, we propose that unambiguous rich semantic descriptions of systems can be developed with well-defined, standardized vocabularies. We focus on the description of relatively well-defined scenarios, rather than attempt to work with unrestricted natural language (also see Allen, in preparation). We build on ontologies (e.g., SUMO; Pease, 2011) and other linguistic resources such as FrameNet (Rupenhoffer, et al., 2016) and VerbNet, which we use to develop descriptions of systems using sequences of transitions. Our models are implemented as object-oriented programs which take advantage of features such as inheritance and concurrency. Indeed, we have proposed an ontology-based semantic UML. There are many advantages to this approach, but challenges remain such as handling granularity and incorporating other constraints.

Ontologies typically provide a hierarchical classification of terms and some of their associated properties. Beyond their specification in ontologies, we need to consider objects in the context of systems. However, many ontologies have limited coverage of processes and the description of the ways that objects typically interact. Unlike objects that fit neatly into hierarchical classification systems, processes have distinct footprints (Ruppenhofer, et al, 2016). The details of processes depend on the objects to which they are applied (compositionality). For instance, the process of opening a door is different from opening a can or opening a restaurant. To handle such differences, and to add tighter constraints in the model, we develop descriptions of object-transition pairs that include the conditions under which the transition will trigger. Object-transition pairs can readily be extended into full propositions with semantic roles as arguments.. Each step adds constraints and still more detail can be added with modifiers for the objects. In addition to expected states, models may include constraints, be broken, or incorporate ambiguous or missing data. A complete corpus of object-transition pairs would be many times larger than current dictionaries but could be developed with semi-automated tools.

Our approach is an alternative to Large Language Models (LLMs). LLMs reportedly often fabricate or "hallucinate" assertions. By comparison, our approach of direct representation provides inspectable and verifiable representations. It supports explainable AI (Mueller, et al., 2021).

## 2.2 Discourse

Even if we have found "natural lines of fracture" for a system, there is leeway in how models are presented. In linguistics, discourse is the study of the intended effects of statements. Typically, discourse refers to interactive communication and includes discourse schemas (McKeown, 1987), discourse macro-structures (Swales, 1990, VanDijk, 1981), and discourse planning (Hovy, 1988). Several types of discourse are usually distinguished: Description, explanation, narrative, negotiation, and argumentation.

Description (or exposition) is subtly distinct from and interlocks with explanation. Explanations are often more didactic. For instance, an explanation might have a forewarning such as "Remember this event, it's important in a later sequence of events". Some aspects are shared across the different types of discourse. Our distinction between description and explanation is related to the distinction between fabula and syuzhet in traditional story-narrative theory. The fabula is the raw events that form the story while syuzhet is the presentation.

There are several accounts of explanation (den Boef & van Woudenberg, 2023; Pitt, 1998). Causation is central in many of them, especially scientific explanations. To distinguish between explanation and narrative it is helpful to consider different senses of narration. In a broad sense, a narrative is any description of connected events, as in narrative history. However, a story narrative is sometimes seen as a distinct type of discourse. An important goal of a story narrative is to manage and manipulate the interest of the reader. For instance, to maintain drama, the outcome of the story is typically undisclosed until the end, whereas in explanation the goal is usually known, even emphasized, from the beginning.

Galileo famously described inertial motion by ignoring friction. The approach of ignoring minor factors has come to be known as Galilean modeling (Thagard, 1999; Weisberg, 2013). In our approach, the causal schema leaves out many details but could allow the user to view them by drilling down. This can be considered an interactive extension of Galilean models.

### 2.3 Causation

In Allen, et al. (2005) we defined causation as a change of state that leads to another change of state. For instance, we say that "sunset causes the temperature to drop". This is a version of the common definition in which a cause is an event *without which* another event would not occur. Causal rules (generalizations or abstractions) also implicitly suggest a comparison to a normative (equilibrium) state. For instance, the assertion smoking causes cancer implies a comparison to the alternative of not smoking. Assertions of causation are often used as a shortcut for presenting a detailed mechanism. However, the assertion does not necessarily imply a detailed understanding of the underlying mechanism (Thagard, 1999). For instance, we may believe that smoking causes cancer without understanding the intermediate steps. There may also be unsubstantiated or only partially confirmed claims of causation but these need to be treated as discourse claims. We can describe sequences of transitions in an equilibrium system, but we do not consider them as causal relationships because, by the definition of an equilibrium system, there are not any system-level state changes. Thus, we can claim that one event is a necessary condition for another event but not that it is a cause (cf., INUS conditions, Mackie, 1974).

## 3 SNOWBALL EARTH EXAMPLE

### 3.1 Overview of Snowball Earth Theory

The Snowball Earth model (Hoffman & Schrag, 2000) is an account of proposed geological processes associated with the development of low-latitude glaciers that froze the surface of the Earth completely. Geologic evidence suggests this happened about 650 million years ago, and possibly a few other times. According to the theory, only when volcanic $CO_2$ in the atmosphere created extreme greenhouse warming did the surface thaw.

### 3.2 Semantic Model of the Snowball Earth Theory

In Allen (in preparation), we developed a small, detailed, state transition description graph for the equilibrium (before any freezing) for part of the theory. One major issue for that was a multilevel representation of streams (collections) of photons. The individual photons had interactions that resulted in collection-level transitions. Here, we consider using chains of state transitions to describe causal relationships. Objects are associated with Dimensions which are properties or processes. Most of those Dimensions are subdivided into States. For the model, states were defined that best illustrated the effects of the transitions; these states could be described more rigorously if needed.

### 3.3 User-Guided Graphical Interaction

We implemented a prototype graphical interface to support causal explanations. It is written in Python with the Tkinter graphics package. The current version is constructed to illustrate and explore key features rather than implement a fully usable service. The curved links are generated with splines but are positioned manually.

Decisions about how to segment and present the model are aspects of discourse. We broke the description of the events into three Episodes (cf., vanDjik, 1981) which are roughly analogous to the chapters in a book. Each episode was composed of an Equilibrium phase that is later disrupted by a Causal (Process) phase: (a) Initial Equilibrium followed by Freezing, (b) Frozen Equilibrium followed by Thaw, and (c) Initial Equilibrium followed by heavy Sedimentation. Note that there are several different types of Sedimentation processes (calcium, iron, magnesium); we will address differentiating these in future work.

### 3.4 System at Equilibrium

Figure 1 shows a screenshot of the interface. At the top right is a widget to select episodes. The control panel is at the lower right. The left (main) panel shows system entities, their dimensions, states, and interaction links between them. It is divided according to sub-regions of the earth. In the upper part of the panel are entities and their dimensions that apply generally across the earth, such as temperature. In the lower portion are transition dimensions associated specifically with the Atmosphere, the Oceans, and the Land.

**Figure 1: Prototype interface with one of the Initial Equilibrium processes in red.**

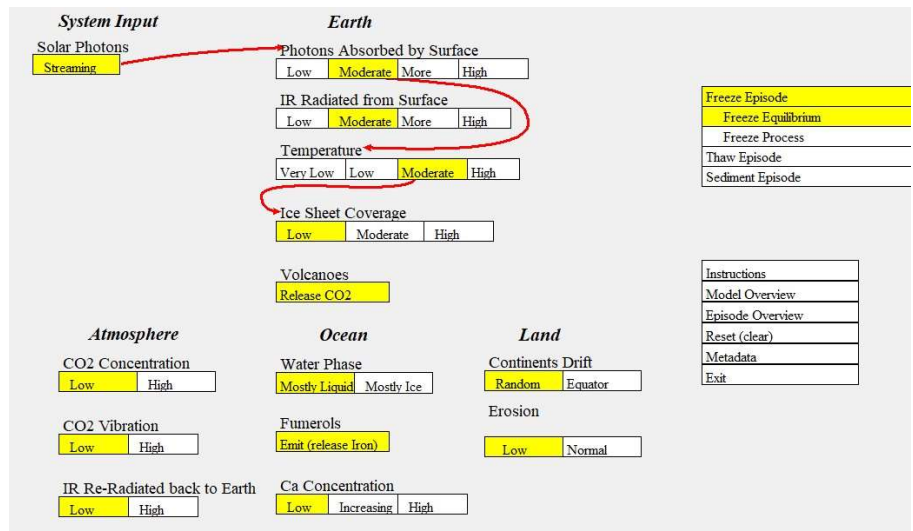

In Figure 1, one of the Freeze Equilibrium processes is shown. Specifically, it shows first that photons from the sun are absorbed into the surface of the earth. The extent of that absorption determines the temperature of the surface, which, in turn, controls the extent of the ice cap. The temperature equilibrium is maintained because an equal amount of heat (Infrared photons) is radiated out from the surface of the earth. This is described in text for the user upon selecting the Episode Overview control panel option. The distinction between visible light photons and IR photons is nuanced. In future work, system-level annotations such as this should be fully structured and incorporated into the knowledge base and graphical presentation.

## 3.5 Causal Processes

Following the Snowball Earth Theory, the system state of primary interest is the ice coverage of the earth's surface. Figure 2 shows the interface for viewing the cause and mechanism associated with extensive freezing over of the earth's surface (bold green arrows). According to the theory, continental drift affected the reflection of photons so much that the earth cooled and freezing was triggered. Specifically, it suggests that the drifting continents aligned near the equator, which increased the albedo, reduced the number of photons absorbed, reduced the temperature, and, thus, increased the ice coverage. As shown in the figure, the equilibrium process (red) is disrupted by the change in reflected light (green) by reducing the number of photons absorbed at the earth's surface.

**Figure 2: As proposed by the theory, the change in the continents' positions disrupts the equilibrium process and causes the earth to freeze over. This image shows the equilibrium process as a red arrow (upper left) and the effect of the drift (green arrows) in the continents' positions on reflected energy, temperature, and ultimately ice cover.**

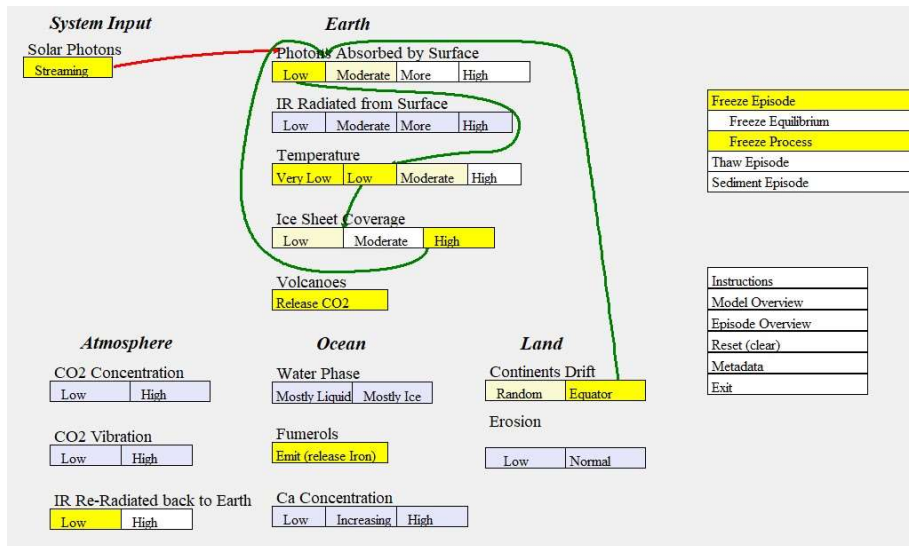

The continental drift transition is shown in the lower center of the figure. The alignment of the continents at the equator is indicated as the change from the light yellow "Random" state to the darker yellow "Equator" state. The causal path (green), leads from there to the photons-absorbed dimension (top). Again, there is a state change with the previous state in light yellow and the later state in yellow. The path continues, resulting in changes in temperature and ice sheet coverage. However, the equilibrium (red) line is terminated since the state of the system has changed. Because the change in the absorption of photons and subsequent changes are due to the alignment of the continents at the equator, we can say that it is the cause of the Snowball freeze.

The final leg of the (green) Freeze Process suggests that the increased ice coverage causes more photons to be reflected (and fewer absorbed) which further reduces the temperature. In other words, there is a positive feedback loop. However, at some point, the supply of liquid water to form the ice will be exhausted and there will be no further increase in the size of the ice cap. In our approach, both the low-level descriptive model and the process threads are sequences of state changes implemented by a program. Potentially the feedback loop could be considered a design pattern.

The change in temperature due to continental drift had other effects and in some cases, the representations overlapped but we do not show those in this view. To avoid ambiguity, most states not associated with the highlighted transitions are grayed out for the presentation of this causal process although they are included in other Episodes.

## 4 Future Work

### 4.1 Incorporating Additional Details

The causal schemas present a relatively high-level overview of the events, based on the underlying structured descriptions (Section 2; Allen, in preparation). Because the schema is abbreviated, users may want to get a deeper understanding by drilling down into the details. For instance, in the schema shown in Figure 1, the reflection of photons by the atmosphere is not mentioned.

Each dimension is associated with an information box that can be accessed by clicking on the label for that the dimension (Figure 3). Potentially, the user could step through the sequence and the descriptions could be presented sequentially. As described in the note, the overall Earth temperature overlaps with the Ocean and Land temperatures. There is a sort of inheritance from the broad earth system to the subsystems (subregions).

**Figure 3. Example of a popup for richer descriptions of the dimensions.**

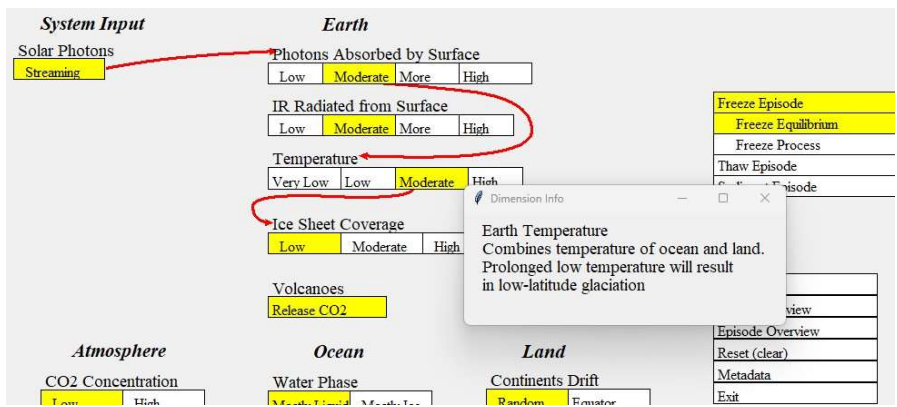

### 4.2 Concurrency, Time, and Timelines

Our model would benefit from a richer representation of time and concurrency. For instance, we noted that the explanation of the Sedimentation processes includes events from both the Freeze and Thaw episodes. The understanding of temporal relationships can be facilitated with graphical timelines (Allen, 2005, 2011).

### 4.3 Multithreaded Narratives

Dynamic systems typically have states and facets that interact in complex, concurrent ways. The situation is analogous to the presentation of a multi-threaded story narrative (e.g., Aronson, 2010). Moreover, many discourses are interactive (e.g., conversations). Maintaining coherence and planning presentations are needed across different types of discourse (e.g., Hovy, 1998).

### 4.4 Verbal Narrative Explanations and User Models

The interface could present an audio narrative (e.g., Allen & Nallaru, 2009) for each of the schemas. Current narratives were simple recountings of the schema transitions. Potentially, they could be much richer and include a range of well-defined discourse moves such as the implications of a given state change for other parts of the system and they could be integrated with the Overview (Section 3) options. Further, the narratives could adapt to the knowledge and interests of the users; adaptive hypertext (Brusilovsky, 1998) and tutoring systems do this. We could also record the propositions the user has viewed and use the record to structure later presentations for that user.

### 4.5 Graphical Animation

User interaction with these system descriptions may be compared to a gamer's interaction with a video game. Techniques for planning for user interaction might be adapted from video games such as Goal Oriented Action Planner (GOAP) and partial order planning (Hartsook, et al., 2011). Perhaps even richer animations could be generated to resemble sequences of scenes analogous to comic strips and full animation could be implemented with graphics engines such as Blender (upbge.org).

### 4.6 Process-based Dynamic Knowledge Graphs

Our work goes beyond typical approaches to knowledge graphs. We include state transitions and higher-level constructs such as workflow mechanisms, complex objects, causal claims, hypotheses, evidence, and even discourse (Allen, 2022; in preparation). Moreover, our knowledge graph would be dynamic in that it is executable and states change as it evolves. Ultimately, we believe that all of this can be coordinated in a single multi-layered knowledge graph. Conflict checks could be implemented with model checking or by using the Python Element Tree to cast the knowledge base into XML.

### 4.7 Relationship to Generative AI

While the modeling in the current approach is laborious, we believe it will become simpler as more tools are developed and the semantic resources are refined. If this work can be scaled, it provides an alternative to the current generation of LLMs. A ChatGPT summary of the Snowball Earth theory we generated is particularly shallow, apparently reflecting the limitations of the current generation of LLM. Nonetheless, LLMs might be useful for generating and refining the knowledge base.

## 5 DISCUSSION

In previous work (Allen, 2022), we explored developing highly structured scientific research reports. This work is related to that because rich, structured descriptions of systems are central to science. We focus on science because scientific presentations are relatively unambiguous, of inherent interest, and have well-developed models. While our earlier work focused on experimental research paradigms, geology is dominated by observation and modeling.

The approach developed here can be extended to incorporate and allow exploration of the evidence for the original Snowball Earth model as well as recent debates about it. Potentially, these rich models could also be extended to describe explanatory coherence and abduction (Thaagard. 1992).  Morevoer, related approaches could be used to develop synthetic languages and models of communities for digital humanities (Allen & Chu, 2021).